\documentclass[12pt]{iopart}
\usepackage{iopams}  
\usepackage{graphicx}
\usepackage{xcolor}
\begin{document}

\title{Quantum Atomic Matter Near Two-Dimensional Materials in Microgravity}

\author{Adrian Del Maestro}
\address{Department of Physics and Astronomy, Min H. Kao Department of Electrical Engineering and Computer Science, and Institute for Advanced Materials and Manufacturing, University of Tennessee, Knoxville, TN 37996}
\ead{Adrian.DelMaestro@utk.edu}

\author{Sang Wook Kim}
\address{Department of Physics and Materials Science Program, University of  Vermont, Burlington, VT 05405}

\author{Nicholas P. Bigelow}
\address{Department of Physics and Astronomy, Institute of Optics, Center for Coherence and Quantum Optics,  University of Rochester, Rochester, NY 14627}

\author{Robert J. Thompson}
\address{Jet Propulsion Laboratory, California Institute of Technology, Pasadena, CA 91109}

\author{Valeri N. Kotov}
\address{Department of Physics and Materials Science Program, University of  Vermont, Burlington, VT 05405}
\ead{Valeri.Kotov@uvm.edu}

\vspace{10pt}

%\begin{indented}
%\item[]May 2022
%\end{indented}

\begin{abstract}
Novel two-dimensional (2D) atomically flat materials, such as graphene and transition-metal dichalcogenides, exhibit unconventional Dirac electronic spectra. We propose to  effectively engineer their interactions with cold atoms in microgravity, leading to a synergy between complex electronic and atomic collective quantum phases and phenomena. 
Dirac materials are susceptible to manipulation and  \emph{quantum engineering} via changes in their electronic properties by application of strain, doping with carriers, adjustment of their dielectric environment, etc. Consequently the interaction of atoms with such materials, namely the van der Waals / Casimir-Polder interaction, can be effectively manipulated, leading to the potential observation of physical effects such as Quantum Reflection off atomically thin materials and confined Bose-Einstein Condensate (BEC) frequency shifts. 
%The exploitation of these effects in furtherance of NASA's fundamental physics mission could result in revolutionary technologies in the fields of energy harvesting, quantum information, atomic sensors, custom film coatings, and materials design. 
\end{abstract}

%\maketitle

\section{Introduction \&  Conceptual Overview  of Relevance to NASA's Mission}

\begin{figure}
\begin{center}
\includegraphics[width=0.4\textwidth]{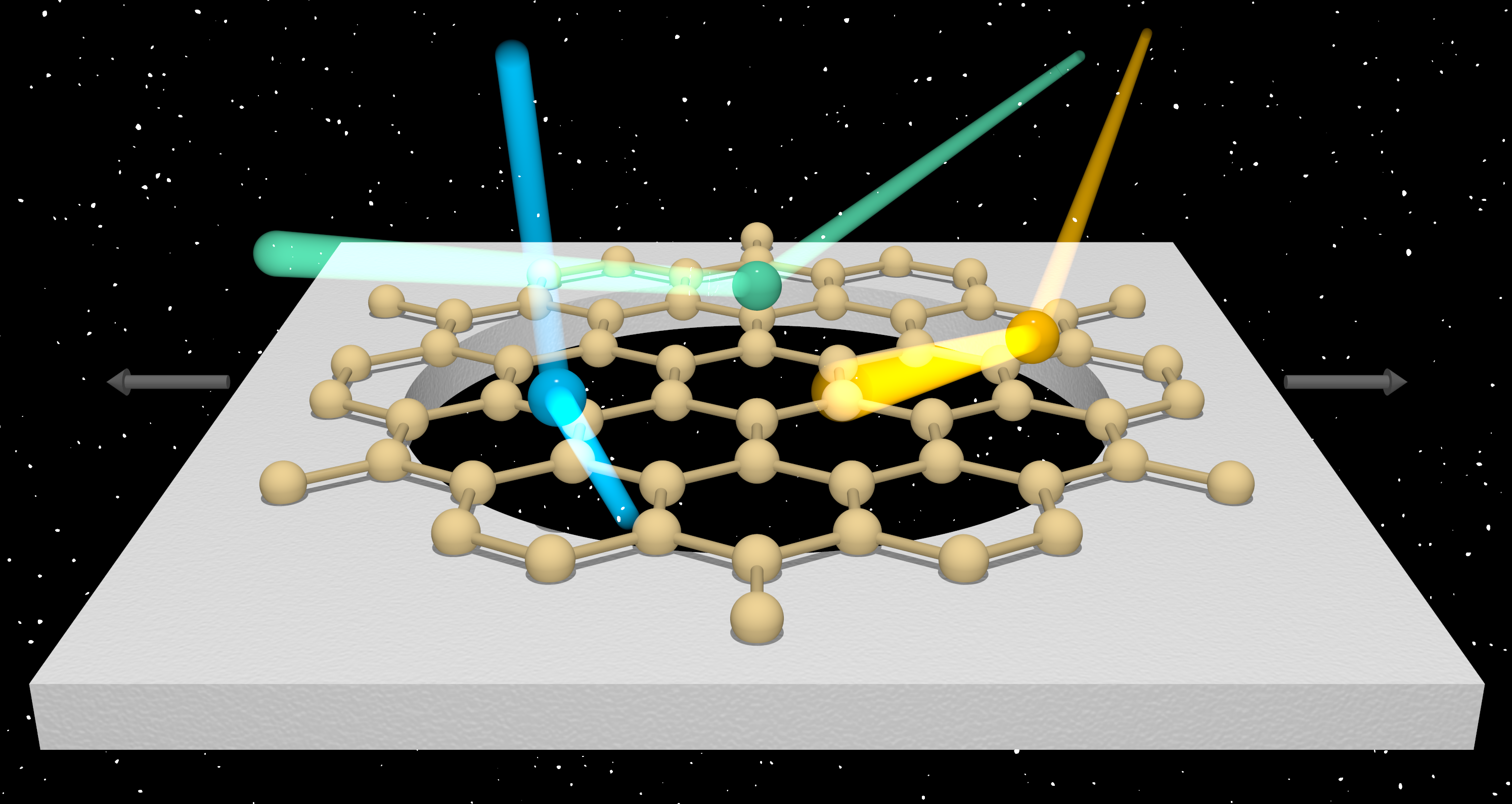}
\includegraphics[width=0.4\textwidth]{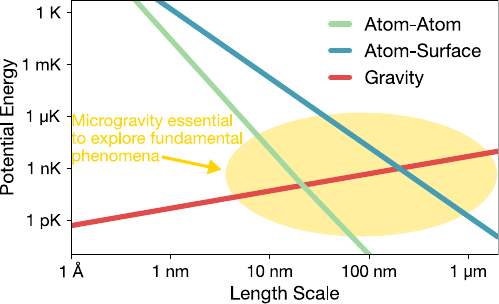}
\end{center}
\caption{\label{fig:atomsMG} Left: Atoms near a two-dimensional free-standing graphene surface in microgravity. Right: A comparison of the energy and length scales where interaction effects between atoms and surfaces may compete with gravity.}
\end{figure}

Novel 2D Dirac materials, which range from semimetals to semiconductors,  form a unique  class of  two-dimensional solids where electrons effectively have a relativistic-like dispersion (massless or massive under certain conditions), with details that are strongly material- and environment-dependent 
\cite{Antonio2009, Geim2013,Manzeli2017,Amorim2016,Kotov2012}. This makes them susceptible to manipulation and  \emph{quantum engineering} by exploring their atomically flat nature and sensitivity of the electronic motion to the lattice structure, electronic density as well as various external factors.  Similar modifications can be realized in other novel 2D materials ``beyond graphene'' \cite{Geim2013,Amorim2016,Naumis2017} widely extending the possibilities for the realization of novel phenomena.
 
 The main driver of unconventional physics is the van der Waals (VDW) / Casimir-Polder (CP) interaction between neutral atoms and 2D Dirac materials \cite{DelMaestro:2021pp,Nichols:2016hd,AVDW-3,AVDW-4,AVDW-5,Henkel:2018}. Here, the unique nature of electron motion causes this interaction to have a well-defined crossover, at the scale of several hundred nanometers,  between the non-relativistic (VDW)  and the relativistic, vacuum fluctuation (Casimir-Polder) components. In general, VDW/CP interactions in atomically flat, graphene-based materials have proven to be of tremendous interest and importance as they  reflect the unique two-dimensional nature of such systems and the inherent possibility for effective manipulation and functionalization 
\cite{vd2,vd3,vd4,vd5,vd6,vd7,vd8,vd9,vd10,Woods:2016,ValeriPhysRevB.87.155431,vdw-anand}.
Therefore this approach offers a unique, materials-based way to study weak dispersion forces which are of fundamental importance in Nature (see Fig.~\ref{fig:atomsMG} where ultracold coherent atoms are released at low momenta near a tunable atomically flat surface), with an impact akin to previous groundbreaking studies on the effects of microgravity on critical phenomena \cite{Barmatz:2007cp}. 

To expose the underlying emergent low energy quantum behavior, the competing interactions and length scales involved may necessitate the microgravity environment of the current and future Cold Atom Laboratory (CAL) missions on the International Space Station. In particular, Figure 1 (right panel) quantifies the interaction potential energy experienced between two light $^4$He atoms (atom-atom) and the van der Waals adsorption energy between one $^4$He atom and a suspended graphene membrane (atom-surface).  Here,  the length scale defines the separation between interacting constituents and the area of interest, depicted by a shaded ellipse, shows where these interactions are within a few orders of magnitude of the gravitational potential energy on earth. For example, based on previous calculations \cite{Nichols:2016hd}, the long range van der Waals adsorption potential becomes comparable to the gravitational potential for helium at distances on the order of $100$~nm above the surface.  While techniques exist for cancelling the effects of gravity in a terrestrial experiment (e.g. through magnetic levitation of atoms), the introduction of a two-dimensional flexible membrane in a suspended, non-strained configuration may necessitate microgravity to prevent sag, which greatly complicates the ability to make detailed predictions connecting theory and experiment. Experiments involving suspended graphene (or other 2D materials), could potentially greatly benefit from ultra low-temperature and microgravity environments 
from purely materials science viewpoint \cite{Ferrari2015, Giustino2020}, in addition to gravity affecting weak  VDW/CP 
interactions as discussed above.  

CAL has already achieved great success in producing trapped Bose-Einstein condensates (BECs) in microgravity \cite{Aveline:2020ie} and it may be possible to explore fundamentally new physical phenomena during the planned BECCAL (Bose-Einstein Condensate Cold Atom Laboratory) mission \cite{frye:epjQuantum_2021} and well beyond, as envisaged by the NASA Fundamental Physics Program. 
This approach represents our vision: to leverage ground and future NASA space-based missions in microgravity for the discovery and engineering of fundamental and exotic physical phenomena at the interface of atomic matter and two-dimensional quantum materials.

The quantum mechanical behavior of nature (statistics, indistinguishability, Heisenberg uncertainty) is clearly on display in two fundamental physical  phenomena: Quantum Reflection of particles above attractive potential tails, with no classical analog,  
  and the formation of  Bose-Einstein condensates (BEC) -- a macroscopically large coherent quantum phase of matter where the participating atoms have settled into their wave-like ground state.  In addition to the BEC's contribution to our understanding of fundamental physics, new discoveries resulting from this research is the foundation for a wealth of applications and new technologies in precision metrology \cite{Fang:2016bq} and quantum information processing \cite{Briegel:2000jj,Henriet:2020rx}. For example, cold atoms have already led to the development of chip-scale atomic clocks with major performance enhancements over legacy technologies \cite{Kitching:2018nh} and the global positioning system is predicated on the precision timing resulting from such atomic clocks.  Consequently, new technological improvements based on cold atom science have the potential to make major and broad-reaching impacts on society. 

Answering fundamental questions related to properties of quantum systems of atoms and molecules, with the aim of designing, for example, sensors using quantum properties,  will require a new generation of experiments, technologies, and  discoveries that necessitate moving to colder temperatures, lower densities, and longer coherence times.  This was broadly recognized by the NASA Fundamental Physics program as detailed in the previous decadal research planning report of the National Research Council: ``Recapturing a Future for Space Exploration, Life and Physical Sciences Research for a New Era'' and renewed during the recent process in 2021. The resulting conceptualization and successful deployment in 2018 of the Cold Atom Laboratory (CAL) \cite{Elliott:2018mg,Aveline:2020ie} now provides the ability to study BECs \cite{Aveline:2020ie} and macroscopic coherent quantum phenomena in the persistent free fall conditions of low Earth orbit. This microgravity environment is exciting and offers the promise of dramatically reducing the forces required to confine ultracold samples of atoms, while simultaneously dealing with the problems of gravitational sag and allowing for long running experiments where the atoms remain nearly fixed relatively to the apparatus.  Long coherence times can be used to do high precision matter wave interferometry experiments  \cite{Becker:2018db,Lachmann:2021mg} while opening up the possibility to study exotic geometries not possible in terrestrial labs, \emph{e.g.}\@ atomic bubbles \cite{Carollo:2022mg}. These recent successes build on a long history of NASA exploiting a microgravity environment to do fundamental physics. This is especially true in the area of universality and critical phenomena \cite{Barmatz:2007cp,Lecoutre:2015iw}, including the most precise confirmation of the existence and value of the anomalous critical exponent of the 3DXY universality class \cite{Lipa:1996eg} which forms the experimental underpinning of the modern renormalization group framework -- one of the most important and successful theoretical methods in physics.

In the rest of the paper we explore two main effects which are quite sensitive to the  VDW/CP interactions with atomically flat quantum materials: (1) quantum reflection of atoms, and (2) trapped Bose-Einstein condensates near 2D materials. These phenomena explore a new set of  ``knobs'' to manipulate cold atoms and BECs in a microgravity environment through the extreme tunability of 2D materials.

%------------------------------------------------------------------------------

\section{Quantum Reflection of Atoms} 

\begin{figure}
\begin{center}
\includegraphics{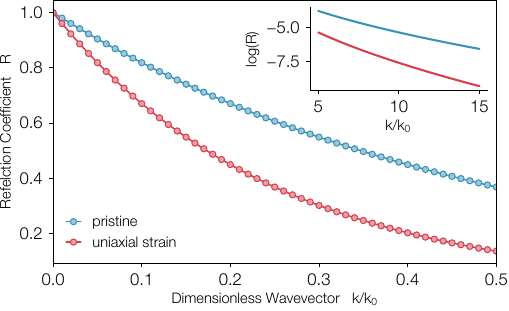} 
\end{center}
\caption{%
\label{fig:reflection-nasa}
The quantum reflection coefficient for Na atoms (with energy $E$ and mass $M$) near pristine and (uniaxially) strained graphene at low atomic wavenumber  $k=\frac{\sqrt{2ME}}{\hbar}$, in units  of  $k_0= 1/\beta_4$, where 
$\beta_4=\frac{\sqrt{2MC_4}}{\hbar}$ \cite{Nichols:2016hd}. The inset shows the high momentum part.}
\end{figure}
One of the most fundamental quantum phenomena with no classical analogue is above-barrier Quantum Reflection (QR).  Scattering of atoms off VDW/CP potential tails \cite{Friedrich2002,Friedrich2004} can provide an extremely sensitive probe of the strength of these fundamental interactions. QR at low atomic energies  has been studied previously for bulk materials, for example by using BECs  \cite{Pasquini2006,Pasquini04} or specular reflection of narrow cold atomic beams \cite{Shimizu1,Shimizu2}.  
Remarkably,  high-energy QR can also be accessed experimentally \cite{dekieviet03}. 
Keeping in mind potential experiments in a microgravity environment, the use of BECs is the most promising direction. 
The low-energy QR in previous  gravitomagnetic  trap experiments tends to saturate below unity due to atomic interaction effects driven by collective excitations of the quantum gas during the reflection \cite{Pasquini2006}. In order to benefit from the unique low atomic velocities achievable with BECs  \cite{deppner:physRevLett_2021}, a release from a shallow trap of  few Hz in necessary. Such traps are, however, heavily distorted by the gravitational pull on Earth, and  recent  observation of BEC in microgravity \cite{Aveline:2020ie} could offer unique opportunities for low-energy QR exploration.

A unique feature of 2D materials, potentially acting as  atomic mirrors, is that accurate theoretical predictions can be made for the VDW interactions and from there the QR, for a variety of materials with different levels of functionalization (i.e.\@ under different  external factors such as strain (Fig.~\ref{fig:reflection-nasa}), carrier density, etc.).  At very low atomic velocities  the quantum reflection tends to the maximum value of unity, and the material surface acts as a perfect atomic mirror.
We also emphasize that as the energy of the atoms decreases the characteristic distance where QR occurs increases \cite{Friedrich2002,Friedrich2004}, meaning that the  weak long-distance tail of the VDW/CP interaction is probed.
 Thus the theory describing low-energy QR, i.e. the potential approach to the atomic mirror limit for different 2D materials,
 is of utmost importance. 
%The atom-surface interactions can be accurately  measured as a phase shift in an atom interferometer where
 %a cold atomic sample with a small drift velocity parallel to a surface is interrogated by four $\pi/4$ pulses such that one branch of the interferometer   can spend long times in the vicinity of a surface of interest.
%These quantum sensors, as planned for BECCAL for example, become extremely sensitive when the drift times are stretched to several seconds as made possible by a microgravity operation.
 
 Let us outline the main  theoretical steps that lead to the QR predictions for a particular 2D material -- graphene. For simplicity, we only write down explicitly  the VDW (non-relativistic part) of the atom-surface interaction potential which has the form:
 
\begin{equation}
\hspace{-1cm} U_{\mbox{vdw}}(z)  =  - \frac{\hbar}{2\pi}\int_{0}^{\infty} d\xi \alpha(i\xi) \    2 \int_{0}^{\infty}dk  k^2 e^{-2kz} 
\int_{0}^{2\pi}
\frac{d\phi}{2\pi} \  \frac{|V(\textbf{k})\Pi(\textbf{k},i\xi)|}{1- V(\textbf{k})\Pi(\textbf{k},i\xi)}.
\label{potential}
\end{equation} 

\noindent
Here $\alpha(i\omega) = \frac{\alpha_0 \omega_0^{2}}{\omega_0^{2}+\omega^{2}}$ is the atomic polarizability
and the relevant parameters for  example for Na atoms are: $\alpha_0= 162.6 \mbox{ a.u.}, \ \omega_0=2.15 \mbox{ eV}$, 
where the atomic unit of polarizability is $1\mbox{ a.u.} = 1.4818 \times 10^{-4} \mbox{ nm}^3$.
The angle $\phi$ is the polar angle measured from the $k_x$  direction. 
$ V(\textbf{k})=2\pi e^2/|\textbf{k}|$ is the Coulomb potential. The polarization function for atomically thin graphene, allowing for strain effects which makes the Dirac cone anisotropic, with different velocities $v_x,v_y$,
is given by the formula (see e.g. \cite{vdw-anand}):

\begin{equation}
\Pi(\textbf{q},i\omega) = -\frac{1}{4 v_{x} v_{y}}\frac{v_{x}^2 q_x^2 + v_{y}^2 q_y^2}{\sqrt{v_{x}^2 q_x^2 + v_{y}^2 q_y^2 +\omega^2}}.
\label{polstrain}
\end{equation} 
%-----
It is well known that for unstrained, isotropic graphene, the value of the velocity is
$v_x=v_y=v = 6.6\ \mbox{eV}${\AA} \cite{Antonio2009}. In the strained case we will assume that
uniaxial strain  $\delta = a/a_0 - 1$, which represents the relative difference between the  lattice spacing  ($a$) in the direction of strain and the unstrained value ($a_0$), is applied in the so-called armchair ($``y"$) direction \cite{Naumis2017}. In order to use the above formulas (\ref{potential}),(\ref{polstrain}), written in momentum space, one needs the modification of the Dirac band structure under strain which  can be extracted from
numerical calculations, summarized in Refs.~\cite{Naumis2017,Nichols:2016hd}.
An excellent fit which works for weak and up to moderate values of strain, is given by:
$v_y/v \approx 1 -2.23 \delta, v_x/v \approx 1 + 2.23 \nu \delta$, where $\nu=0.165$ is the Poisson ratio for graphene. Naturally the quasiparticle velocity decreases in the strain direction while exhibiting a slight increase in the perpendicular direction. For example for strain 10\% ($\delta = 0.1$) we have Dirac cone anisotropy
$v_y/v_x \approx 0.75$, while increasing strain up to  34\%, we obtain $v_y/v_x \approx 0.2$.

It should be noted that  the non-relativistic form,   Eq.~(\ref{potential}), in itself contains a crossover  
from $-C_3/z^3$ to  $-C_4/z^4$
behavior in the potential (with log corrections) which reflects the motion of Dirac quasiparticles in graphene.
Relativistic corrections are then taken into account as described in \cite{Nichols:2016hd,AVDW-4,Henkel:2018};
we do not present the corresponding fully relativistic formulas. While we use the full expressions, in practice, for
distances up to $\sim100$ nm, we find that the relativistic effects are quite small and become gradually more  pronounced only as the distance increases. 

The calculation of the QR coefficient $R$ for Na follows the methodology outlined in \cite{Friedrich2002}.
For this atom our calculations \cite{Nichols:2016hd} show that the $-C_4/z^4$  tail is dominant and the potential is fitted to the form  $U(z) = -\frac{C_4}{z^4} \equiv -  \frac{\hbar^2}{2M}\frac{\beta_4^2}{z^4}$ which defines the length scale 
$\beta_4=\frac{\sqrt{2MC_4}}{\hbar}$. Then in the asymptotic limits of low and high atomic energies one finds the behavior
\cite{Friedrich2002}:

\begin{equation}
R \approx 1 - 2 (\beta_4 k) ,  \ \ \beta_4 k \ll 1 \ ; \ \ \ \ \ \  R \sim  e^{-1.694  \sqrt{\beta_4 k}}, \  \  \  \beta_4 k \gg 1.
\label{QR}
\end{equation}

Figure~\ref{fig:reflection-nasa} shows theoretical predictions for  suspended pristine graphene and uniaxially strained
sample (corresponding, for illustration purposes, to fairly large strain  34\% ($v_y/v_x = 0.2$)). Since strain leads to larger graphene polarizability and thus enhanced attractive potential, it decreases QR relative to pristine graphene. 
The opposite tendency, i.e. an increase of QR at given energy is expected to take place  when physical factors that lower 
the 2D material polarizability are at play. For example this can occur in relatively small  graphene samples which exhibit an effective finite-size electronic gap, or in 2D materials, such as dichalcogenides, with an intrinsic gap.

Because the  phenomenon of Quantum Reflection is highly sensitive to the electronic motion in atomically thin materials and can   
in principle be accurately calculated in a variety of physical situations, and for different atoms and 2D materials, we believe 
it offers a promising route towards characterization of weak VDW/CP forces, especially in a microgravity environment.

%------------------------------------------------------------------------------

\section{Trapped BECs near 2D Materials as Ultrasensitive Force Sensors}

\begin{figure}
\begin{center}
\includegraphics[width=0.45\textwidth]{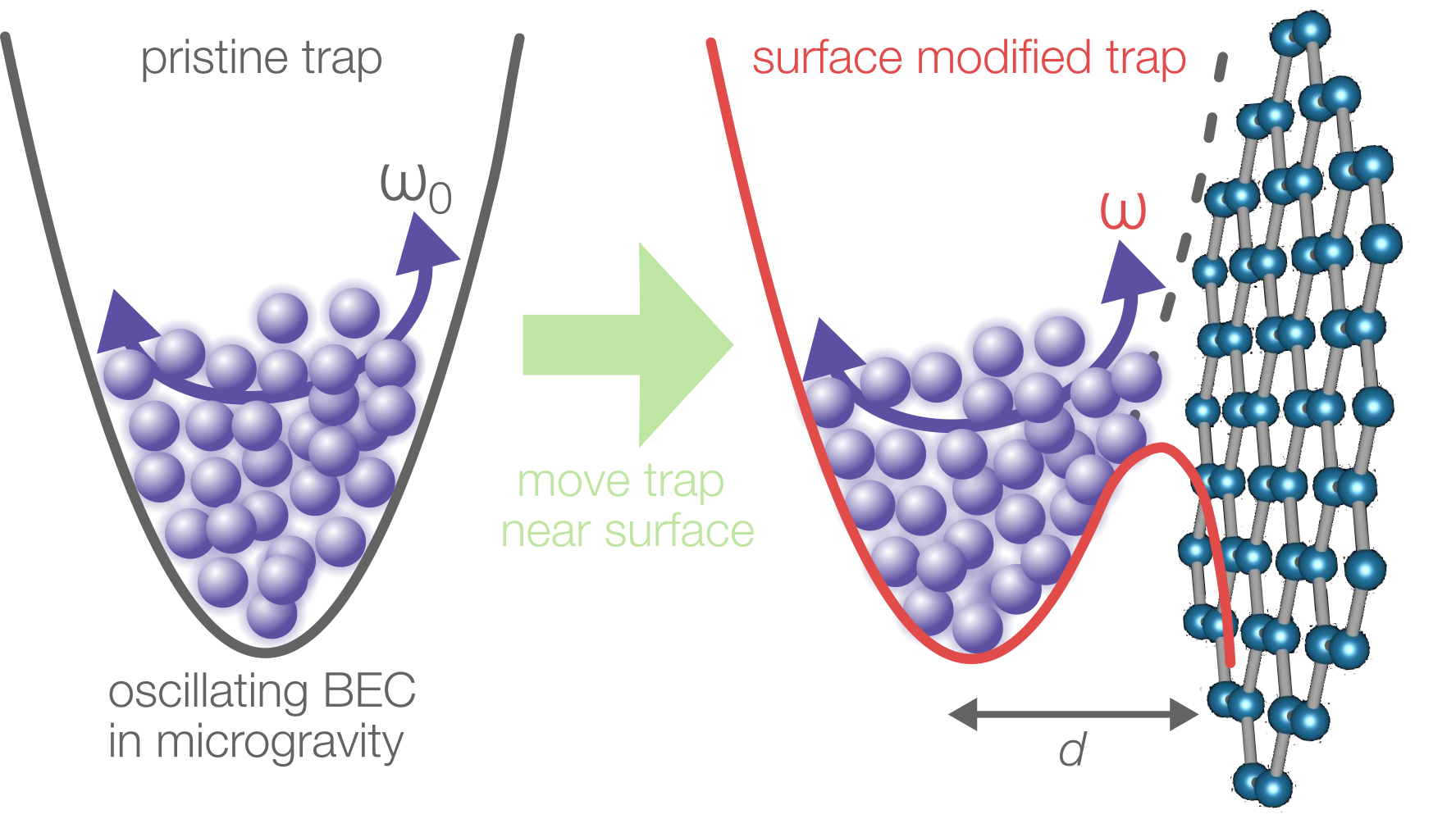} 
\includegraphics[width=0.45\textwidth]{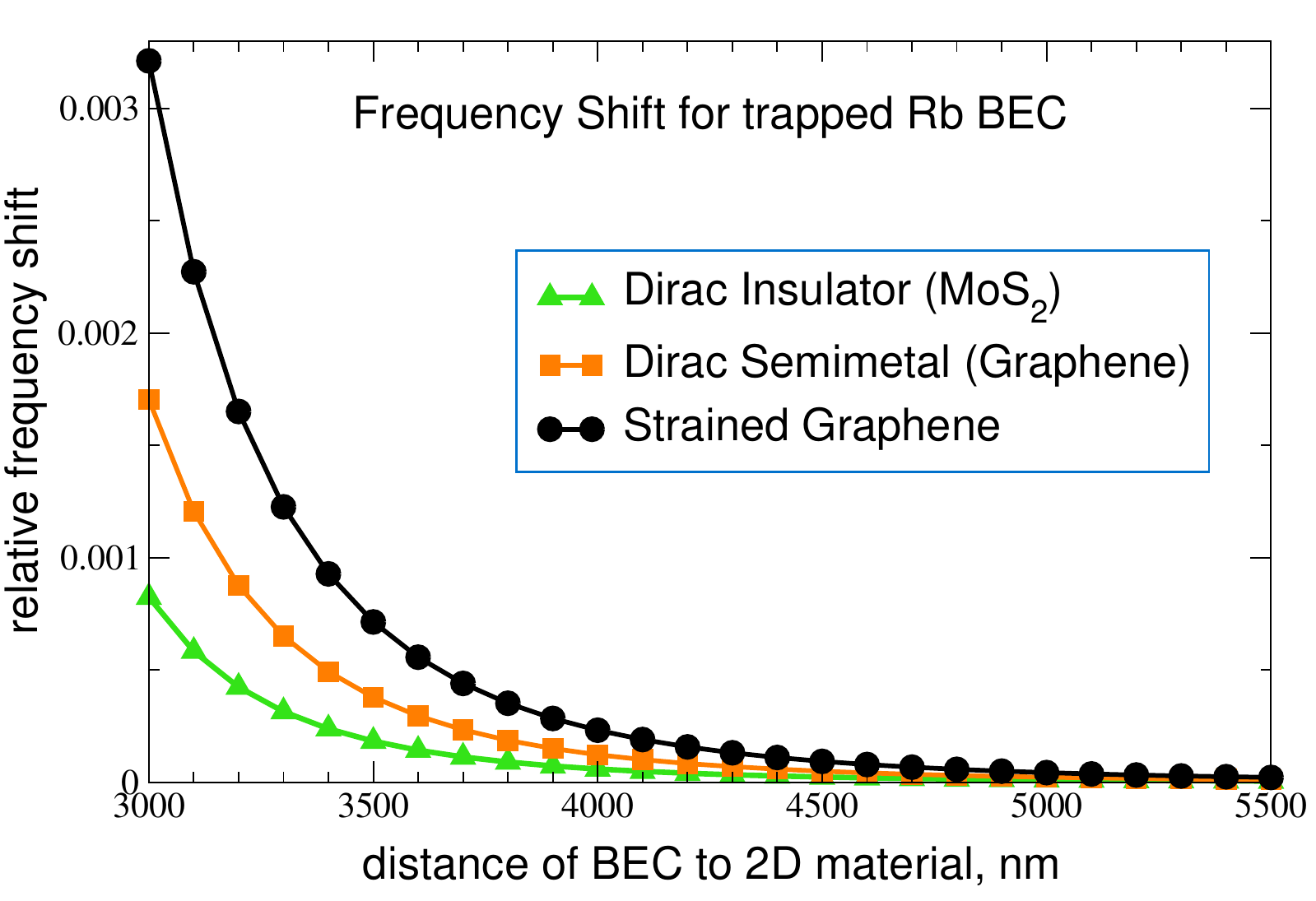} 
\end{center}
\caption{%
\label{fig:BEC}
Bose-Einstein Condensates near 2D surfaces. Left: Schematic diagram of confined BEC near material showing the modification of the trapping potential. Right: Calculated relative change of the frequency of center of mass oscillations, $(\omega_0-\omega)/\omega_0$, for different 2D materials, versus distance $d$ to the surface.}
\end{figure}

The study of trapped BECs near material surfaces offers another opportunity to utilize the unique capabilities of BECCAL and a microgravity environment for the manipulation and thus precision measurement of the  VDW/CP force. The cold atom trapping potential is modified due to the attractive force of the material atoms which can result in a noticeable change in the BEC condensate's center of mass oscillation frequency \cite{Antezza:2004,Cornell2005} which is protected from gravitational sagging effects (see Fig.~\ref{fig:BEC}).  Advances in atom-on-chip techniques \cite{Keil2016,GrapheneAChips}, and in particular the utilization of 2D materials such as graphene,  will make it possible to place the BEC even closer to the material (of order hundreds of nanometers),  without suffering any losses and maintaining high atom lifetimes. Experiments in microgravity can produce ultra-coherent condensates \cite{Aveline:2020ie} necessary to push the boundaries of quantum force measurement.

We have performed calculations for Rb condensates near graphene (pristine and strained) and MoS$_2$ (2D insulator).
The atom-material potential $U(z)$ in the case of graphene is calculated following the procedure 
described in the previous section.
The parameters characterizing the polarizability of Rb are:  $\alpha_0= 318.6 \mbox{ a.u.}, \ \omega_0=1.67\ \mbox{eV}$.
The polarization function of MoS$_2$, which is characterized by massive Dirac quasiparticles (i.e. an electronic gap)
can be taken as:

\begin{equation}
\hspace{-2.3cm}   
\Pi(\textbf{q}, i\omega) = -\frac{|\textbf{q}|^{2}}{\pi}\bigg[\frac{m}{\tilde{q}^{2}} + \frac{1}{2\tilde{q}}\bigg(1-\frac{4m^{2}}{\tilde{q}^{2}}\bigg)\tan^{-1}\bigg(\frac{\tilde{q}}{2m}\bigg)\bigg],
     \    \tilde{q} \equiv \sqrt{v^2|\textbf{q}|^{2} + \omega^2},    \  m = \Delta/2.
\label{polgap}     
\end{equation}
The  values for MoS$_2$ are: $\Delta =  1.66 \ \mbox{eV}$, $v= 3.51\ \mbox{eV}${\AA}.
A summary of material parameters for  members of the dichalcogenide family can be found in Refs.~\cite{Manzeli2017,wetting}.

The center of mass frequency $\omega$ is smaller than the  value $\omega_0$ (determined by the parameters of the BEC confining potential, harmonic trap) due to the attractive nature of the VDW/CP potential between the 
condensate atoms and the surface.
Following the approach of Ref.~\cite{Antezza:2004},  the equation governing the behavior of $\omega$ is:

\begin{equation}
\omega^2-\omega_0^2 = \frac{1}{M}\int_{-R_z}^{R_z}n_{0} (z)\frac{\partial^2 U(z)}{\partial z^2}dz,
\label{freqshift}
\end{equation}

\noindent
where $M$ is the mass of the Rb atom and $n_{0}(z)$  is the effective  BEC density along the $z$ direction 
(1D ``column" density) as described in \cite{Antezza:2004,Dalfovo:1999}. The quantity $R_z \approx 2 \ \mu\mbox{m}$ is the Thomas-Fermi 
radius of the condensate.

Figure~\ref{fig:BEC} shows  the  calculated frequency change as a function of the distance $d$ between the 2D material and the center of the harmonic trap. One can  take the optimistic viewpoint that  the BEC can maintain its structure  much closer to the surface (than in previous bulk cases) as suggested in \cite{GrapheneAChips}. Our most important observations are: (1) At a given distance the frequency change is about an order or magnitude smaller than for bulk samples \cite{Cornell2005}, as expected for atomically thin configuration. It is still within the limits of experimental detection. (2) The frequency change is very sensitive to the type of material since it reflects the change in 2D polarization properties (in the plot, strain makes graphene more polarizable while MoS$_2$ is a gapped Dirac material and thus less polarizable).

To conclude, theoretically-predicted frequency changes of BECs near 2D Dirac materials in microgravity show extraordinary sensitivity to material parameters and therefore this set-up can be used as a powerful and ultrasensitive probe of the nature and strength of VDW/CP interactions.

\section{Outlook: Beyond Conventional Materials Science $\rightarrow$ Functional Intelligent Materials}
The phenomena discussed in this paper are in principle possible within the present level of our theoretical understanding and current NASA technologies.  Looking beyond the near-term we can explore some exciting and transformational trends. It is quite possible that the above effects can be further ``designed'' and engineered, in the following sense. Given the wide variety of currently known 2D materials, it is potentially feasible to construct materials, using artificial intelligence \cite{Carleo:2019ml}  with specific properties, optimized in such a way that their interaction with atoms has the correct strength for a given desired functionality.  This could provide unprecedented control over the fundamental van der Waals / Casimir-Polder force, never previously achieved in a theoretical or laboratory setting. The NASA fundamental physics program can play a decisive role in this process through its unique ability to provide an accessible microgravity laboratory able to probe the quantum effects of atoms near 2D materials.

%\cite{McGuirk04, Vuletic04}

\section{Acknowledgements}
This work was supported, in part, under NASA grant number 80NSSC19M0143. A portion of this research was carried out at the Jet Propulsion Laboratory, California Institute of Technology, under a contract with the National Aeronautics and Space Administration.
\\

\bibliographystyle{iopart-num.bst}
\bibliography{references.bib} 

\end{document}